\documentclass[11pt]{article}
\pdfoutput=1 
\usepackage{a4wide}
\usepackage{epsfig,epsf,graphics}
\usepackage{times}
\usepackage{color}
\usepackage{amsfonts,amssymb,latexsym,amsmath} 
\usepackage{cleveref}
\usepackage{bm}
\usepackage{enumerate}
\usepackage{enumitem}
\usepackage{hhline}
\usepackage{longtable}
\usepackage{array} 
\usepackage[
colorlinks=true,
linkcolor=black,
breaklinks=true,
urlcolor=green,
citecolor=blue]{hyperref}
\usepackage{cite}
\usepackage{epstopdf}

\newcommand{\al}{&\!\!\!\!}

\newcommand{\Amp}{\mathcal{A}}
\newcommand{\Tr}{\textrm{Tr}}

\newcommand{\order}[1]{\mathcal{O}\left(#1\right)}
\renewcommand{\arraystretch}{1.2}

\newcommand{\email}[1]{\footnote{{\em E-mail address:} \texttt{#1}}}

\begin{document}

\title{Hindered magnetic dipole transitions  between $P$-wave
bottomonia and coupled-channel effects}

\author{Feng-Kun Guo$^{\, a,}$\email{fkguo@itp.ac.cn} ,
Ulf-G.~Mei{\ss}ner$^{\, a,b,c,}$\email{meissner@hiskp.uni-bonn.de} ,
Zhi Yang$^{\, a,b,}$\email{zhiyang@hiskp.uni-bonn.de}\\ 
    {\it\small$^a$CAS Key Laboratory of Theoretical Physics, Institute of
      Theoretical Physics,}\\
      {\it\small  Chinese Academy of Sciences, Beijing 100190, China}\\
    {\it\small$^b$Helmholtz-Institut f\"ur Strahlen- und Kernphysik and Bethe
   Center for Theoretical Physics, }\\
   {\it\small Universit\"at Bonn,  D-53115 Bonn, Germany }\\
   {\it\small$^c$Institut f\"{u}r Kernphysik, Institute for Advanced
Simulation, and J\"ulich Center for Hadron
Physics,}\\
   {\it \small Forschungszentrum J\"ulich, D-52425 J\"{u}lich, Germany}
   }

\maketitle

\begin{abstract}

\noindent In the hindered magnetic dipole transitions of heavy quarkonia, the
coupled-channel effects originating from the coupling of quarkonia to a pair of heavy
and anti-heavy mesons can play a dominant role. Here, we study the hindered magnetic
dipole transitions between two $P$-wave bottomonia, $\chi_b(n P)$ and
$h_b(n^\prime P)$, with $n\neq n^\prime$.
In these processes the coupled-channel effects are expected to lead to partial
widths much larger than the quark model predictions.
We estimate these partial widths which, however, are very sensitive to unknown
coupling constants related to  the vertices $\chi_{b0}(nP)B\bar B$.
A measurement of the hindered M1 transitions can shed light on the
coupled-channel dynamics in these transitions and hence on the size of the
coupling constants. We also suggest to check the coupled-channel effects by
comparing results from quenched and fully dynamical lattice QCD calculations.

\end{abstract}

\thispagestyle{empty}

\newpage

In recent years, several new bottomonia were discovered. One of the most
interesting discoveries is the $h_{b}(1P)$ found in the puzzling $\pi^0$
transition $\Upsilon(3S) \to \pi^0 h_{b}(1P)$ with a subsequent electric dipole
(E1) transition to the $\eta_b(1S)$ by the Babar
collaboration~\cite{Lees:2011zp}. This finding is consistent with the prediction
that such a transition is a promising way to produce the
$h_b$~\cite{Voloshin:1985em,Godfrey:2002rp}.
The isospin violating decay channel has the same final states, $\gamma\gamma
h_{b}$, as the one in the electromagnetic cascades $\Upsilon(3S)\to \gamma
\chi_{bJ}(2P)\,(J=0,1,2)$ and $\chi_{bJ}(2P)\to \gamma h_b$. The branching
fractions for the E1 transitions $\Upsilon(3S)\to \gamma \chi_{bJ}(2P)$ are well
measured to be of the order of 10\%, but no experimental result for the hindered
magnetic dipole (M1) transition $\chi_{bJ}(2P)\to\gamma h_b$ is available.
Thus, it is important to investigate the decay channel $\chi_{bJ}(2P)\to\gamma
h_b$.
The $h_b(1P)$ later on was also observed in the isospin conserving decay process
$\Upsilon(4S)\to\eta h_b$~\cite{Tamponi:2015xzb} with a branching fraction
$(2.18\pm0.21)\times10^{-3}$, consistent with the estimate of the order
$10^{-3}$ in Ref.~\cite{Guo:2010ca}, where this channel was suggested to be used
to search for the $h_b$.

The quark model has been used to study the spectrum and decay properties of the
excited bottomonia without the coupled-channel effects from intermediate
open-bottom mesons~\cite{Godfrey:2015dia}. The spectrum was also calculated with
the inclusion of coupled-channel effects~\cite{Liu:2011yp}. More generally, we
remark that coupled-channel effects due the virtual hadronic loops are of recent
interest in heavy quarkonium physics.
In the quenched quark model, the mixture between the bare hadron states and the
two-meson continuum is not taken into account. When the coupled-channel effects
are considered, the quarkonium spectrum gets shifted~(the values of these mass
shifts depend on the specific models, see, e.g.,
Refs.~\cite{Eichten:1979ms,Heikkila:1983wd,Eichten:2005ga,Pennington:2007xr,Barnes:2007xu,Li:2009ad,Danilkin:2009hr,
Liu:2011yp,Ferretti:2012zz,Zhou:2013ada}).
In addition to the impact on the mass spectrum, the coupled-channel effects are
expected to be important in some transitions between heavy
quarkonia~\cite{Lipkin:1988tg,Moxhay:1988ri,
Zhou:1990ik,Meng:2007tk,Meng:2008bq,Simonov:2008qy,Guo:2009wr,Guo:2010ak,
Li:2011ssa,Mehen:2011tp,Chen:2011jp}.
In particular, they are expected to dominate the hindered M1 transitions between
the $P$-wave quarkonia because of two reasons: first, the hindered M1
transitions break heavy quark spin symmetry and their widths in the quark model
come from relativistic corrections; second, the coupled-channel contribution has
an enhancement due to the $S$-wave couplings of the two vertices involving heavy
quarkonia~\cite{Guo:2011dv}.
For instance, the partial width of $\chi_{c2}(2P) \to \gamma h_c(1P)$ from the
coupled-channel effects is two orders of magnitude larger than the prediction
from the quark model as shown in Ref.~\cite{Guo:2011dv}.
Such hindered M1 transitions between bottomonia may be measured at
Belle-II~\cite{roberto}. However, although there have been calculations on
hindered M1 transitions between $S$-wave heavy quarkonia in the framework of
effective field theory~\cite{Brambilla:2005zw,Pineda:2013lta} and lattice
QCD~\cite{Lewis:2011ti,Lewis:2012bh,Becirevic:2014rda,Hughes:2015dba}, so far
only a few predictions on similar transitions between $P$-wave bottomonia have
been given, and all of them are based on quark model
calculations~\cite{Godfrey:2015dia}. Since in bottomonium systems the
relativistic corrections are small, the quark model predictions on these partial
widths are tiny, in the range from sub-eV to eV. Yet, similar to the charmonia
case, the coupled-channel effects due to virtual bottom mesons could enhance the
decay widths to values that make an observation possible. This motivates us to
study here the hindered M1 transitions between $P$-wave bottomonia by
considering the coupled-channel effects through  coupling to virtual bottom and
anti-bottom mesons. An additional important motivation for us to study these
processes is the fact that experimentalists plan to study them at the coming
Belle-II experiment~\cite{roberto}.

\begin{figure}[t!]
\begin{center}
  \includegraphics[width=0.4\textwidth]{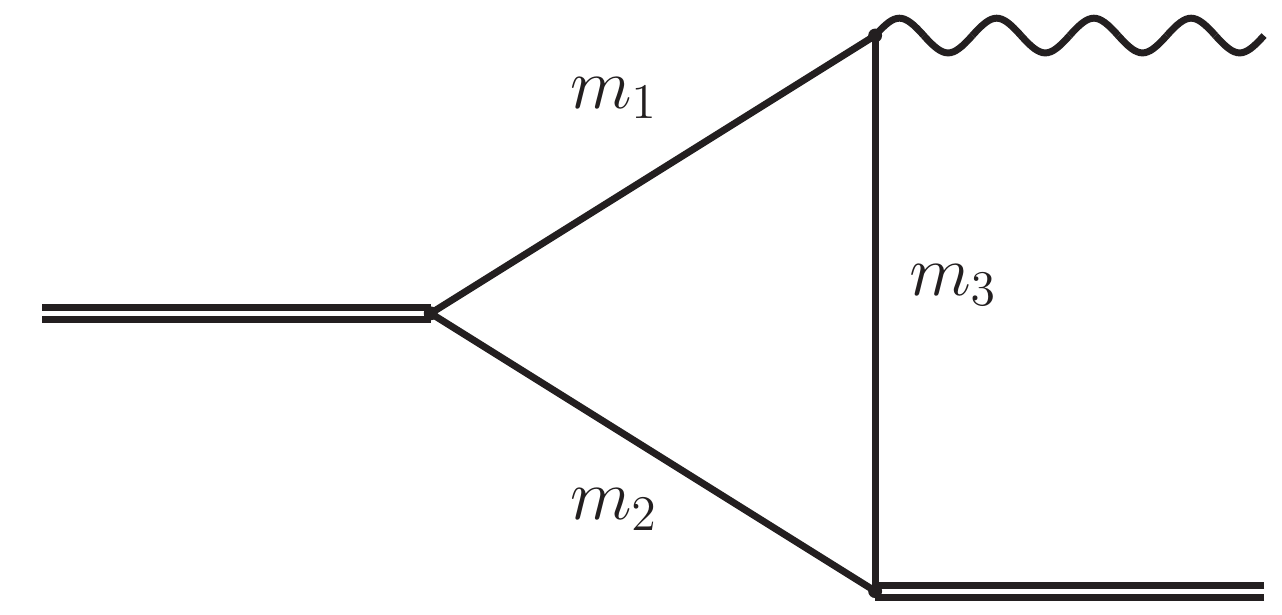}
\caption{Triangle diagram where the double, solid and wavy lines
represent the bottomonium, bottomed meson and the photon, respectively.}
\label{fig:LoopDiagram}
\end{center}
\end{figure}

Due to the fact that the bottomonia are close to the open bottom thresholds so
that the intermediate bottom mesons are nonrelativistic, we use 
nonrelativistic effective field theory (NREFT) suitable for investigating such
coupled-channel effects in heavy quarkonia
transitions~\cite{Guo:2010ak,Guo:2009wr,Guo:2010zk}.
The intermediate mesons are nonrelativistic so that their velocities, denoted by
$v$, are much smaller than one, and the loop diagrams scale in powers of $v$.
The three-momentum and kinetic energy are counted as $v$ and $v^2$,
respectively, and each of the nonrelativistic propagators scales as $v^{-2}$.
Further, a $P$-wave bottomonium couples to a pair of ground state bottom and
anti-bottom mesons in an $S$-wave. At leading order, the coupling is described
by a constant which does not contribute any power to the velocity counting.
Thus, the triangle diagram in Fig.~\ref{fig:LoopDiagram} scales
as~\cite{Guo:2011dv}
\begin{equation}
 \Amp_\text{triangle} \propto
 \frac{v^5}{(v^2)^3}\frac{E_{\gamma}}{m_b}=\frac{E_{\gamma}}{v m_b},
 \label{eq:triangle}
\end{equation}
where the factors $1/m_b$ and $E_\gamma$ are due to the spin-flip of the heavy
quark in M1 transitions and the $P$-wave coupling of the photon to the
bottom mesons, respectively. One thus sees that the closer the bottomnia to the
bottom-meson thresholds, the larger the coupled-channel effects. One remark is
in order: $v$ in the power counting is in fact the average of two velocities.
This can be estimated as $v=(v_i+v_f)/2$ with $v_i=\sqrt{|m_1+m_2-M_i|/\bar
m_{12}}$ and $v_f=\sqrt{|m_2+m_3-M_f|/\bar m_{23}}$, where $m_{1,2,3}$ are the
masses of intermediate mesons as labelled in Fig.~\ref{fig:LoopDiagram},
$M_{i(f)}$ is the mass for the initial (final) bottomonium, and $\bar m_{jk}$
is the averaged value of $m_j$ and $m_k$.

\begin{figure}[tb]
\begin{center}
  \includegraphics[width=0.8\textwidth]{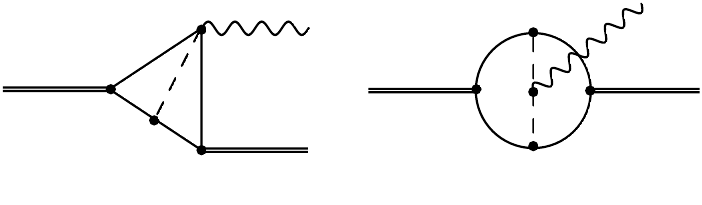}
\caption{Two typical two-loop diagrams where the double, solid and wavy lines
are the same as in Fig.~\ref{fig:LoopDiagram} and the dashed lines represent
the exchanged pions.}
\label{fig:2loop}
\end{center}
\end{figure}

However, unlike the case of charmonium hindered M1 transitions, the two-loop
diagrams with a pion exchanged between two intermediate bottom mesons are not
highly suppressed for the bottomonium transitions. From the power counting
analysis in Ref.~\cite{Guo:2011dv}, the relative importance of the two-loop
diagrams shown in Fig.~\ref{fig:2loop} in comparison with the triangle diagram
given in Fig.~\ref{fig:LoopDiagram} can be described by a factor
\begin{eqnarray}
  \label{eq:2loop}
  \frac{\Amp_\text{2-loop}}{\Amp_\text{triangle}} \sim v \frac{g^2
  M_B^2}{\Lambda_{\chi}^2} ,
\end{eqnarray}
where $M_B$ is the bottom
meson mass, $\Lambda_{\chi}=4\pi F_\pi$, with $F_\pi$  the pion decay
constant, is the chiral symmetry breaking scale, and $g\simeq0.5$ is the axial
coupling constant for bottom mesons~\cite{Detmold:2011bp,Bernardoni:2014kla,
Flynn:2015xna}. This ratio can be understood as follows (taking the left
diagram in Fig.~\ref{fig:2loop} as an example): the two
more propagators and one more nonrelativistic loop integral measure, in
comparison with the diagram in Fig.~\ref{fig:LoopDiagram}, together give the
factor $v=v^5/(v^2)^2$ in the above equation; $g^2/\Lambda_\chi^2$ comes from
the two pionic vertices and one more loop; $M_B^2$ is introduced to make the
ratio dimensionless. Taking the masses of the $1P$, $2P$ and $3P$ bottomonia
from Refs.~\cite{Aaij:2014hla,Agashe:2014kda}, the velocity in the power
counting may be estimated to be 0.31, 0.23 and 0.18 for the $2P\to1P$, $3P\to1P$
and $3P\to2P$ radiative transitions, respectively. One then finds that the
relative factor given in Eq.~\eqref{eq:2loop} is of order one, which means that
the contribution of two-loop diagrams like the ones shown in
Fig.~\ref{fig:2loop} should be of similar size as the one-loop triangle diagram
in Fig.~\ref{fig:LoopDiagram}. This is different from the charmonium case
studied in Ref.~\cite{Guo:2011dv} where $M_B^2$ is replaced by the much smaller
$M_D^2$ and thus leads to a suppression. Nevertheless, we will only calculate
the triangle diagram, and keep in mind that given the power counting of the
two-loop diagrams such a calculation can only be regarded as an estimate, rather
than a precise calculation, with a quantitative uncertainty analysis out of
reach.

As a result of the approximate heavy quark spin symmetry, one can classify the
heavy-light bottom mesons according to the total angular momentum of the light
degrees of freedom $s_\ell$ and collect them in doublets with total spin
$J=s_\ell\pm\frac{1}{2}$. For instance, the pseudoscalar ($P_a$) and vector
($V_a$) bottom mesons are collected in the spin multiplet with
$s^P_\ell=\frac{1}{2}^-$.
The two-component effective fields~\cite{Hu:2005gf} that describe the ground
state heavy mesons in the heavy quark limit are $H_a=\vec{V}_a\cdot
\vec{\sigma}+ P_a$ for annihilating bottom mesons and $\bar H_a
=-\vec{\bar V}_a\cdot \vec{\sigma}+\bar P_a$ for annihilating anti-bottom mesons, where
$\vec{\sigma}$ are the Pauli matrices and $a$ is the light flavor index.
Moreover, the $P$-wave bottomonia can be collected in a spin multiplet as
\begin{equation}
\chi^i=\sigma^j\left(-\chi^{ij}_{b2}-\frac{1}{\sqrt{2}}\epsilon^{ijk}\chi^{k}_{b1}
+\frac{1}{\sqrt{3}}\delta^{ij}\chi_{b0}\right)+h^i_b\,.
\end{equation}

As mentioned above, the leading order coupling of the $P$-wave bottomonium to
the bottom and anti-bottom mesons is in an $S$-wave, and thus is
given by~\cite{Colangelo:2003sa,Guo:2010zk}
\begin{equation}
\mathcal{L}_{\chi}=i\frac{g_1}{2} \Tr [\chi^{\dag i} H_a \sigma^i
\bar{H}_a]+h.c.,
\end{equation}
where $\Tr$ denotes the trace in the spinor space.
We also need the magnetic coupling of the photon to the $S$-wave heavy
mesons~\cite{Amundson:1992yp,Hu:2005gf,Mehen:2011tp}
\begin{eqnarray}
  \label{eq:lagD}
  \mathcal{L_\gamma} = \frac{e\,\beta}{2} \Tr\left[ H_a^\dag H_b^{}\,
\vec{\sigma}\cdot \vec{B} \, Q_{ab}^{} \right] + \frac{e\, Q'}{2m_Q} \Tr \left[
H_a^\dag \,\vec{\sigma}\cdot \vec{B} \, H_a^{}\right],
\end{eqnarray}
where $B^k=\epsilon^{ijk}\partial^iA^j$ is the magnetic field,
$Q_{ab}=\text{diag}(2/3,-1/3,-1/3)$ is the light quark electric charge matrix,
$Q'$ is the heavy quark electric charge (for a bottom quark, $Q'=-1/3$ ), and
$m_Q$ is the mass of the heavy quark.

\medskip

\begin{table*}[t]
\begin{center}
\renewcommand{\arraystretch}{1.3}
\begin{tabular}{|l|l|}\hline\hline
$\chi_{b0} \to h_b\gamma$ & $[B^*,{\bar B}^*,B]$, $[B^*,{\bar B}^*,B^*]$,
 $[B,{\bar B},B^*]$ \\
$\chi_{b1} \to h_b\gamma$ & $[B^*,{\bar B},B^*]$, $[B,{\bar B}^*,B^*]$\\
$\chi_{b2} \to h_b\gamma$ & $[B^*,{\bar B}^*,B]$, $[B^*,{\bar B}^*,B^*]$\\
$h_{b} \to \chi_{b0}\gamma$ & $[B^*,{\bar B},B]$, $[B,{\bar	B}^*,B^*]$, $[B^*,{\bar B}^*,B^*]$\\
$h_{b} \to \chi_{b1}\gamma$ & $[B^*,{\bar B},B^*]$,
$[B^*,{\bar B}^*,B]$\\
$h_{b} \to \chi_{b2}\gamma$ & $[B,{\bar B}^*,B^*]$, $[B^*,{\bar	B}^*,B^*]$\\
\hline\hline%
\end{tabular}
\caption{Triangle loops contributing to each transition, where the mesons are
listed as $[m1,m2,m3]$ corresponding to the notations in
Fig.~\ref{fig:LoopDiagram}. For simplicity, the charge
conjugation modes and the light flavor labels are not shown here.
}
\label{tab:loops}
\end{center}
\end{table*}

We specify the intermediate mesons in the list $[m_1,m_2,m_3]$, as denoted in
Fig.~\ref{fig:LoopDiagram}. All the possible loops with the intermediate
pseudoscalar and vector bottomed mesons are listed in Table~\ref{tab:loops} for
the corresponding transitions. The pertinent transition amplitudes are given in the
appendix.
From these amplitudes, one clearly sees two sources of spin symmetry breaking:
the terms from the bottom quark magnetic moment are explicitly proportional to
$1/m_b$, and the sum of $\beta$-terms in each amplitude vanishes if the vector
and pseudoscalar bottom mesons are taken to be degenerate.\footnote{For
Eqs.~(\ref{eq:chib0hb},\ref{eq:chib1hb},\ref{eq:hbchib0},\ref{eq:hbchib1}) given in the
appendix, this point is apparent, for Eqs.~(\ref{eq:chib2hb},\ref{eq:hbchib2}), one can
see this after taking the absolute value squared of the amplitude and summing up
the polarizations.}

The loops involved here are convergent, which means that the coupled-channel
effects for the processes of interest are dominated by long-distance physics
described in our NREFT. We do not need to introduce a counterterm here. The
situation is different for the case of E1 transitions. The loop integrals
involved there are divergent, and thus the contact term considered in
Ref.~\cite{Mehen:2011tp} also serves as a counterterm and is necessary for
renormalization.

Using the masses of  the mesons given by the Particle Data
Group~\cite{Agashe:2014kda}, it is easy to get numerical results for the partial
decay widths. As for the masses of the $3P$ bottomonia, we choose the quark model
values from Ref.~\cite{Godfrey:2015dia}, which were obtained based on the
measured $\chi_{bJ}(3P)$ mass by the LHCb Collaboration~\cite{Aaij:2014hla} with
the predicted multiplet mass splittings, i.e. $M_{h_b(3P)}=10.519$~GeV,
$M_{\chi_{b0}(3P)}=10.500$~GeV, $M_{\chi_{b1}(3P)}=10.518$~GeV and
$M_{\chi_{b2}(3P)}=10.528$~GeV. These masses are very close to the ones in
Ref.~\cite{Liu:2011yp}, where the coupled-channel effects are taken into account
in a nonrelativistic quark model. We also take
$\beta=1/276$~MeV$^{-1}$~\cite{Hu:2005gf}, and $m_b=4.9$~GeV.

\begin{table}[th]
\begin{center}
\begin{tabular}{| l | c  c  c | c |} \hline\hline
                                     & $J=0$ & $J=1$ & $J=2$  & units \\
 \hline
 $\chi_{bJ}(3P) \to h_b(2P)\gamma$   & 0.3  & 1.8  & 1.4   & $(g_1^\prime
 g_1^{\prime\prime})^2$ keV   \\%
 $h_{b}(3P) \to \chi_{bJ}(2P)\gamma$ & 0.3  & 2.2  & 1.6   & $(g_1^\prime
 g_1^{\prime\prime})^2$ keV   \\%
 \hline
 $\chi_{bJ}(3P) \to h_b(1P)\gamma$   & 4.9 & 13.4 & 11.9   & $(g_1
 g_1^{\prime\prime})^2$ keV  \\%
 $h_{b}(3P) \to \chi_{bJ}(1P)\gamma$ & 3.3  & 15.8 & 15.4   & $(g_1
 g_1^{\prime\prime})^2$ keV  \\%
 \hline
 $\chi_{bJ}(2P) \to h_b(1P)\gamma$   & 1.2  & 1.8  & 1.8   & $(g_1
 g_1^\prime)^2$ keV   \\%
 $h_{b}(2P) \to \chi_{bJ}(1P)\gamma$ & 0.7  & 2.0  & 2.5   & $(g_1
 g_1^\prime)^2$ keV   \\%
\hline \hline
\end{tabular}
\caption{Decay widths for the hindered M1 transitions between $\chi_{bJ}(nP)$
and $h_b(n^\prime P)$, where the coupling constants take values in units of
$\rm{GeV}^{-1/2}$.
}
\label{tab:widths1}
\end{center}
\end{table}
The decay amplitudes are proportional to the product squared of the coupling
constants of the bottom and anti-bottom mesons to the $1P$, $2P$ and $3P$ bottomonia, 
denoted as $g_1$, $g_1^\prime$ and
$g_1^{\prime\prime}$, respectively. As the mass of the $\chi_{bJ}(1P,2P,3P)$ and
$h_{b}(1P,2P,3P)$ are below the bottom and anti-bottom meson threshold, the
coupling constants cannot be measured directly. Here, we show the decay width
of the hindered M1 transitions between two $P$-wave bottomonia in units of
the coupling constants in the Table~\ref{tab:widths1}.
%
\begin{table}[th]
\begin{center}
\renewcommand{\arraystretch}{1.4}
\begin{tabular}{| l | c | c | c | c | c | c | c |} \hline\hline
&\multicolumn{2}{|c|}{$J=0$} & \multicolumn{2}{|c|}{$J=1$} & \multicolumn{2}{|c|}{$J=2$} \\
\hline
& ours & RQM & ours & RQM & ours & RQM  \\
\hline
$\displaystyle\frac{\Gamma_{h_{b}(2P) \to \chi_{bJ}(1P)\gamma}}{\Gamma_{\chi_{bJ}(2P) \to
h_b(1P)\gamma}}$ & $0.59$ & $0.03$ & $1.1$ & $0.5$ & $1.4$ & $9.2$ \\%
\hline \hline
\end{tabular}
\caption{Comparison of the ratios of the decay
widths for the $2P$ to $1P$ bottomonia with the ones from the
RQM~\cite{Godfrey:2015dia}.}
\label{tab:widthsratio}
\end{center}
\end{table}
%
The unknown parameters will get cancelled if we calculate ratios of the
decay widths which are proportional to the same product squared of coupling
constants. Furthermore, we also expect that these ratios are less sensitive to
the two-loop diagrams in Fig.~\ref{fig:2loop} as the numerator and denominator
in the ratio, being related to each other via spin symmetry, would get  a
similar correction.
The ratios in our calculation can be easily obtained from Table~\ref{tab:widths1}. 
In order to show that the
coupled-channel effects lead to very different values for some of these ratios,
we show a comparison of ratios for selected decay widths of the hindered M1
transitions between the $2P$ to $1P$ bottomonia with those obtained in the
quenched quark model of Ref.~\cite{Godfrey:2015dia} in
Table~\ref{tab:widthsratio}.
These predictions can be tested in the future from experiments or
lattice QCD calculations. In fact, radiative transitions of $S$-wave
bottomonia, including the hindered M1 ones, have been studied by using
lattice QCD~\cite{Lewis:2011ti,Lewis:2012bh,Hughes:2015dba}. As suggested in
Ref.~\cite{Guo:2011dv}, one can check the coupled-channel effects directly in
lattice QCD by comparing results in full and quenched calculations ---
the former includes the coupled-channel effects intrinsically while the latter
does not.

As mentioned in Ref.~\cite{Godfrey:2015dia}, the numerical results of these
hindered transitions in the quark model are very sensitive to relativistic
corrections (these transitions do not vanish only when relativistic corrections are
accounted for in quenched quark model). Nevertheless, they are tiny because the
M1 transitions break heavy quark spin symmetry as well, and are in the ballpark of
sub-eV to eV in Ref.~\cite{Godfrey:2015dia}. If the partial widths really take
such small values, an experimental observation of the bottomonium hindered M1
transitions would be impossible in the foreseeable future. In turn, this means
that once such transitions are observed, the mechanism would be different from
that in the quenched quark model, and would be caused by  coupled-channel
effects. Then, the measured partial widths can be used to estimate the involved
coupling constants.

Unfortunately, the values of the coupling constants $g_1,g_1'$ and $g_1''$
cannot be estimated reliably. If one takes the model estimate made in
Ref.~\cite{Colangelo:2003sa},\footnote{Here we have replaced the charmonium
quantities by the corresponding bottomonium ones, and there is a factor of 2
difference for $g_1$ in the definition of the Lagrangian in \eqref{eq:lagD} and
that in Ref.~\cite{Colangelo:2003sa}.}
$g_1=-2\sqrt{m_{\chi_{b0}}/3}/f_{\chi_{b0}}$ and uses the value
$f_{\chi_{b0}}\approx175$~MeV from a QCD sum rule
calculation~\cite{Azizi:2012mp}, then one gets $g_1\sim-20$~GeV$^{-1/2}$. This
value is so large that if the $\chi_{b0}$ is located only 1~MeV above the
$B^0\bar B^0$ threshold it would have a huge width of 21~GeV. However, the quark
model predictions for the open-bottom partial decay widths of the $4P$
bottomonia leads to $|g_1(4P)|\sim0.2$~GeV$^{-1/2}$ (the one for the $5P$ states
is slightly smaller), which, although it is for the $4P$ states, is two orders
of magnitude smaller than that from the former estimate.
In Ref.~\cite{Guo:2011dv}, the product of the coupling constants $(g_1
g_1^\prime)^2$ is estimated to be of order ${\cal O}(10~\text{GeV}^{-2})$ in the
charm sector, where the difference between the model estimate for
$g_1$~\cite{Colangelo:2003sa} and the extracted value from quark model
predictions of the $2P$ charmonium decay widths is much smaller.
If we naively take the same estimate here, despite that there is no simple
flavor symmetry between charmonia and bottomonia, then the partial decay widths of
$\order{1\sim10^2}$~keV could be large enough for a possible measurement in the
future.

In principle, we can also calculate the decay widths for the isospin breaking
transitions between the $\chi_{bJ}(nP)$ states with the emission of one pion.
They would be proportional to the same combination of unknown coupling
constants. The charmonium analogues from the coupled-channel
effects have been analyzed in details in Ref.~\cite{Guo:2010ak}.  However, we
refrain from such a calculation because the isospin breaking between the
charged and neutral bottom mesons is one order of magnitude smaller than that in
the charmed sector because of the destructive interference between the
contributions from the up and down quark mass difference and the
electromagnetic effect~\cite{Guo:2008ns}.

In summary, we studied the hindered M1 transitions between two $P$-wave
bottomonia, $\chi_b(n P)$ and $h_b(n^\prime P)$ ($n\neq n^\prime$) assuming the
mechanism is dominated by coupled-channel effects. Because of the suppression
from heavy quark spin breaking and small relativistic corrections, such
transitions have tiny partial widths from sub-eV to eV in quark model. In the
mechanism underlying coupled-channel effects, the breaking of heavy quark spin
symmetry can come from the different masses of bottom mesons within the same
spin multiplet, and the problem of tiny matrix elements for transitions between
bottomonia of different principal quantum numbers in the quark model does not exist
as well.
Therefore, it is natural to expect that the coupled-channel effects lead to much
larger widths for such transitions than those predicted in the quark model. A future
observation of such transitions at, e.g., Belle-II~\cite{roberto} may be
regarded as a clear signal of the coupled-channel effects, and the measured
widths could then be used to extract a rough value of the product of the so-far
unknown coupling constants, e.g. $g_1 g_1'$. Such information would be useful
for other transitions where intermediate bottom mesons play an important role,
such as the decays of the $Z_b(10610)$ and $Z_b(10650)$ into $h_b\pi$ and
$h_b(2P)\pi$.

At last, we want to emphasize again that the coupled-channel effects in heavy
quarkonium transitions can be checked directly in lattice QCD by comparing results from quenched
and fully dynamical simulations as we already suggested in Ref.~\cite{Guo:2011dv}.
A better understanding of coupled-channel effects would lead to new insights
into the dynamics of heavy quarkonia.

\bigskip

\section*{Acknowledgments}

We would like to thank Roberto Mussa for discussions and encouraging us to
perform this study during the 2nd B2TiP Workshop. Two of the authors (UGM, ZY)
gratefully acknowledge the hospitality at the ITP where this work was performed.
This work is supported in part
by the DFG and the NSFC through funds provided to the Sino-German CRC~110
``Symmetries and the Emergence of Structure in QCD'' (NSFC Grant No.
11261130311), by the Thousand Talents Plan for Young Professionals, and by the
Chinese Academy of Sciences President's International Fellowship Initiative
(Grant No. 2015VMA076).

\bigskip

\begin{appendix}

\section{Decay amplitudes}

The decay amplitude for each diagram is the sum of all possible triangle
diagrams, and each diagram can be expressed in terms of convergent scalar
three-point loop functions~\cite{Guo:2010ak}
\begin{eqnarray}
\label{eq:loop1}
I(m_1,m_2,m_3) \al=\al i\int\!\frac{d^4l}{(2\pi)^4}
\frac{1}{\left(l^2-m_1^2+i\epsilon\right)
\left[(P-l)^2-m_2^2+i\epsilon\right]
\left[(l-q)^2-m_3^2+i\epsilon\right] } \nonumber\\
\al=\al \frac{\mu_{12}\mu_{23}}{16\pi m_1m_2m_3} \frac{1}{\sqrt{a}}
\left[ \tan^{-1}\left(\frac{c'-c}{2\sqrt{ac}}\right) +
\tan^{-1}\left(\frac{2a+c-c'}{2\sqrt{a(c'-a)}}\right) \right].
\end{eqnarray}
where $a = (\mu_{23}/m_3)^2 \vec{ q}\ ^2$, $c =
2\mu_{12}b_{12}$, $c'=2\mu_{23}b_{23}+(\mu_{23}/m_3)\vec{q}\ ^2$,
$\mu_{ij}=m_im_j/(m_i+m_j)$, $b_{12} =
m_1+m_2-M$ and $b_{23}=m_2+m_3+q^0-M$. In the loop function, $P$ and $q$ are the
momenta of the initial bottomium and the photon, respectively, and
$m_{i}(i=1,2,3)$ are the mass of the intermediate mesons. In the deriving of
Eq.~(\ref{eq:loop1}), the nonrelativistic approximation has been
adopted.

The pertinent amplitudes for the decays are listed here:
\begin{eqnarray}
 \mathcal{M}_{\chi_{b0}\to\gamma h_b}\al=\al
-\frac{2 i e g g^{\prime}}{\sqrt{3}} q^i \varepsilon
^j(\gamma)\epsilon _{ijk} \varepsilon ^k(h_b)\sum_{a=u,d,s}  \bigg\{2
\left(\beta Q_a +\frac1{3 m_b}\right) I(B^*_a,\bar{B}^*_a,B^{*}_a)\nonumber\\
\al\al + \left(\beta Q_a -\frac1{3m_b}\right)
\left[ I(B^{*}_a,\bar{B}^{*}_a,B_a) - 3I(B_a,\bar{B}_a,B^{*}_a) \right]
\bigg\}, \label{eq:chib0hb}\\
 \mathcal{M}_{\chi_{b1}\to\gamma h_b}\al=\al
{2 i \sqrt{2} e g g^{\prime}}
\left[\vec q\cdot \vec\varepsilon (\chi_{b1}) \vec\varepsilon
   (\gamma)\cdot \vec\varepsilon (h_b)-\vec q\cdot \vec\varepsilon
   (h_b) \vec\varepsilon (\gamma)\cdot \vec\varepsilon
   (\chi_{b1})\right]\nonumber\\
\al\al \times \sum_{a=u,d,s} \left[ \left(\beta  Q_a + \frac1{3m_b}\right)
   I(B^*_a,\bar{B}_a,B^*_a)
  - \left(\beta  Q_a - \frac1{3m_b}\right) I(B_a,\bar{B}^{*}_a,B^{*}_a) \right]
   , \label{eq:chib1hb}\\
 \mathcal{M}_{\chi_{b2}\to\gamma h_b}\al=\al
4 i e g g^{\prime}\epsilon _{ijk} \varepsilon ^{kl}(\chi_{b2})\sum_{a=u,d,s}
\bigg\{
- q^i \varepsilon ^j(\gamma)  \varepsilon
  ^l(h_b) \left(\beta Q_a -\frac1{3m_b}\right) I(B^{{*}}_a,\bar{B}^{{*}}_a,B_a)
   \nonumber\\
  \al\al +\varepsilon ^i(h_b) \left[q^l \varepsilon
^j(\gamma) - q^j \varepsilon ^l(\gamma)\right]
\left(\beta  Q_a + \frac1{3m_b}\right) I(B^{{*}}_a,\bar{B}^{{*}}_a,B^{{*}}_a)
  \bigg\},
  \label{eq:chib2hb}\\
 \mathcal{M}_{h_b\to \gamma\chi_{b0}}\al=\al
-\frac{2 i e g g^{\prime}}{\sqrt{3} } q^i \varepsilon ^j(\gamma)
\varepsilon ^k(h_b) \epsilon _{ijk} \sum_{a=u,d,s} \bigg\{2
\left(\beta Q_a +\frac1{3m_b}\right) I(B_a^{{*}},\bar{B}^{{*}}_a,B^{{*}}_a)
\nonumber\\
   \al\al  + \left(\beta  Q_a - \frac1{3 m_b}\right)
   \left[I(B_a,\bar{B}^{{*}}_a,B^{{*}}_a) - 3I(B^{{*}}_a,\bar{B}_a,B_a)\right]
   \bigg\}, \label{eq:hbchib0}\\
 \mathcal{M}_{h_b\to \gamma\chi_{b1}}\al=\al
 {2 i \sqrt{2} e g g^{\prime}}
\left[\vec q\cdot \vec\varepsilon (\chi_{b1}) \vec\varepsilon
   (\gamma)\cdot \vec\varepsilon (h_b)-\vec q\cdot \vec\varepsilon
   (h_b) \vec\varepsilon (\gamma)\cdot \vec\varepsilon
   (\chi_{b1})\right]\nonumber\\
\al\al \times \sum_{a=u,d,s} \left[ \left(\beta  Q_a + \frac1{3m_b}\right)
   I(B^*_a,\bar{B}_a,B^*_a)
  - \left(\beta  Q_a - \frac1{3m_b}\right) I(B^{*}_a,\bar{B}^{*}_a,B_a) \right],
 \label{eq:hbchib1}\\
 \mathcal{M}_{h_b\to \gamma\chi_{b2}}\al=\al
4 i e g g^{\prime} \epsilon _{ijk} \varepsilon
^{kl}(\chi_{b2}) \sum_{a=u,d,s}\bigg\{
- q^i \varepsilon ^j(\gamma) \varepsilon ^l(h_b) \left(\beta  Q_a -
   \frac1{3 m_b}\right) I(B_a,\bar{B}^{{*}}_a,B^{{*}}_a)
\nonumber\\
   \al\al  \varepsilon ^i(h_b) \left[q^l \varepsilon
^j(\gamma) - q^j \varepsilon ^l(\gamma)\right]
\left(\beta  Q_a + \frac1{3m_b}\right) I(B^{{*}}_a,\bar{B}^{{*}}_a,B^{{*}}_a)
\bigg\},
   \label{eq:hbchib2}
\end{eqnarray}
where the initial bottomonium should be understood to be of higher excitation
then the final one, $\varepsilon^i(\gamma)$, $\varepsilon^i(h_b)$ and $\varepsilon^i(\chi_{b1})$ are the polarization vectors for the photon, $h_b$
and $\chi_{b1}$, respectively, and $\varepsilon^{ij}(\chi_{b2})$ is the
symmetric polarization tensor for the $\chi_{b2}$. One also needs to notice that
a factor $\sqrt{M_iM_f}$, with $M_{i,f}$ denoting the masses of the initial
and final bottomonia, should be multiplied to each of the amplitudes to account
for the nonrelativistic normalizations of the heavy quarkonium fields (similar factors for the intermediate heavy mesons have been obsorbed in the
definition of the loop function).

\end{appendix}

\end{document}